\documentstyle[eqsecnum,multicol,aps,psfig]{revtex}

\draft

\headheight=\baselineskip
\begin{document}
 
\bibliographystyle{prsty}


\title{On the global hydration kinetics of tricalcium silicate cement 
}

\author{F. Tzschichholz$^{1,2}$ and H. Zanni$^2$}
 
\address{
$^1$ Department of Physics, 
Norwegian University of Science and Technology,\\ 
N--7034 Trondheim, Norway}
\smallskip
\address{
$^2$ Laboratoire de Physique et M\'ecanique des Milieux H\'et\'erog\`enes
(CNRS, URA 857),\\ \'Ecole Sup\'erieure de Physique 
et de Chimie Industrielle de la Ville de Paris,\\
10 rue Vauquelin, 75231 Paris Cedex 05, France}
 
\date{\today}
\maketitle
 
\begin{abstract}
We reconsider a number of measurements for the overall
hydration kinetics of tricalcium silicate pastes having an initial water 
to cement weight ratio close to $0.5$. We find that the time dependent 
ratio of hydrated and unhydrated silica mole numbers can be 
well characterized by two power--laws 
in time, $x/(1-x)\sim (t/t_\times)^\psi$. For early times 
$t < t_\times$ we find an `accelerated' hydration 
($\psi = 5/2$) and for later times $t> t_\times$ 
a `deaccelerated' behavior ($\psi = 1/2$). The crossover 
time is estimated as $t_\times \approx 16\;hours$.
We interpret these results in terms of a global second order rate equation
indicating that (a) hydrates catalyse the hydration
process for $t<t_\times$, (b) they inhibit further hydration for
$t>t_\times$ and (c) the value of the associated second order rate 
constant is of magnitude 
$6\cdot 10^{-7}\, -\,7\cdot 10^{-6}\;liter\;mol^{-1}\;s^{-1}$.
We argue, by considering the hydration process actually being furnished as 
a diffusion limited precipitation that the exponents 
$\psi =5/2$ and $\psi =1/2$ directly indicate a preferentially 
`plate' like hydrate microstructure. This is essentially in agreement 
with experimental observations of cellular hydrate microstructures for
this class of materials.
 \end{abstract}
 
%

\newsavebox{\boxl}
\savebox{\boxl}{$\ell$}
 
\begin{multicols}{2}
\section{Introduction}

\bigskip
Heterogeneous solid--state transformations are of great importance 
and practical interest 
in industrial and technological applications.
Perhaps most prominent examples are transformations within alloys
and steels playing a so crucial role for their mechanical and
durabillity properties. 
Solid--state transformations are frequently accompanied by 
microstructural changes within the material in form of  
precipitating (segregating) phases from (solid) solutions. In its
simplest form the transformation takes place between a major component
acting as a solvent and an initially solvated phase. 
The precipitation process of the solvated phase is usually considered 
to happen in three 
main stages; phase nucleation, phase growth, 
and finally a coarsening process leading to a homogenisation of 
microstructural geometrical properties\cite{Hann75,Saut96}. Further below 
we will employ basic concepts of solid--state transformation in order to 
interpret the experimentally observed hydration kinetics of cement, 
See Section \ref{Interpretation}.     

The massive introduction of hydraulic binders into daily engineering 
problems challenges a better understanding on how these materials work.
Beside this are cementing processes also of great importance for 
certain geological problems, i.e., for the formation, properties and behaviour
of sedimentary bassins\cite{Oelk96}.
Typical hydraulic binders are plaster, cement, mortar, and concrete. 
In the following we will focus on the hydration of neat Portland 
cement (tricalcium silicate) more closely. 
   
Basic phenomenological aspects of cement hydration 
can be characterized as follows. 
Initially fine grained cement powder    
(here tricalcium silicate, $C_3S \equiv Ca_3SiO_5$, with grain 
diameters typically ranging between 
$5\,\mu m$ and $50\, \mu m$\cite{c1}) 
is mixed well with water. 
Tricalcium silicate is a well crystallized 
compound ( $\rho = 3.21\,g\,cm^{-3}$)
and the employed powders typically have specific surface areas
of order of 
$3\cdot 10^3 \,cm^2\,g^{-1}$\cite{b3}. 
Rapidly after mixing the cement particles start to
dissolve. The principal reaction 
products are solvated ions ($Ca^{2+}$, $OH^{-}$ and $H_2SiO_4^{2-}$) 
diffusing within the solvent. 
The ion concentrations are bounded themselve by finite 
solubility products above which hydrate phases start 
to precipitate from solution, preferable on surfaces of
already existing hydrates.
There are two associated hydrates, a) the cement hydrate 
($C_{1.5}SH_{2.5} \equiv (CaO)_{1.5}\,(SiO_2)\,(H_2O)_{2.5}$) and b)
the Portlandite ($CH \equiv Ca(OH)_2 $), which compete for the  
common calcium and hydroxyl ions.
The cement hydrate ( $\rho = 2.35\, g\,cm^{-3}$) is amorphous in
contrast to the crystalline Portlandite ( $\rho = 2.24\, g\,cm^{-3}$). 

The process of cement dissolution, ion diffusion, and hydrate 
precipitation is usually referred as `cement hydration'. 
The ion diffusion represents the physical coupling between the
chemical dissolution and precipitation reactions 
leading to a complex physico--chemical evolution of 
hydrate microstructure. 

\begin{figure}[htb]
\centerline{\psfig{file=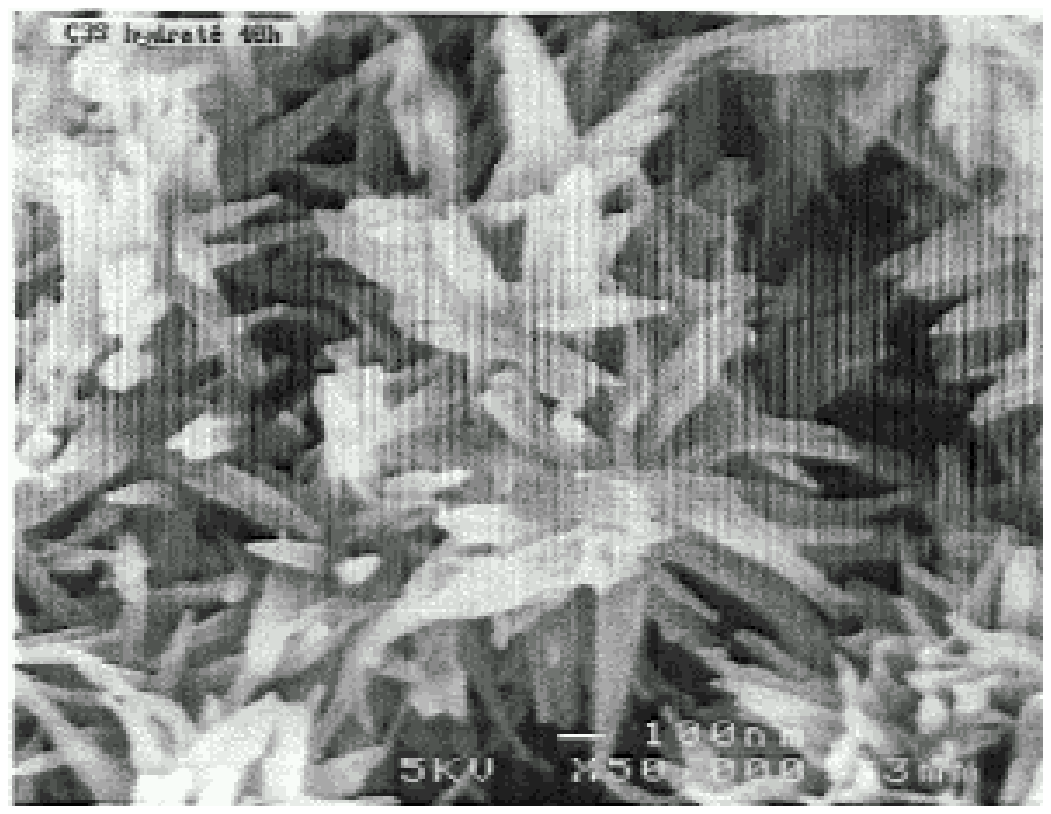,width=\textwidth}}
\vskip 5mm
\caption{
SEM micrograph of hydrated tricalcium silicate
cement after $40$ hours hydration time. The 
initial water to cement ratio was $0.5$. Note the leaf--like hydrate 
microstructure (with courtesy of Institute Francais du Petrole).
  }
\label{fig0}
\end{figure}

In Fig.~\ref{fig0} we show a hydrated 
microstructure after $40$ hours of hydration time. Typical is the 
leaf/foil--like structure.

In the following we shall propose on the basis of already published
NMR measurements an 
empirical global reaction rate law for the hydration of cement pastes 
for initial water to cement weight ratios $w/c =0.5$. The observed 
values for the kinetics exponents can be understood -- in parts -- by 
classical solid--state transformation theory.

\section{Experimental Results and Kinetic Relationships}
The hydration of cement is spontaneous (exothermic) and irreversible
and can be characterized by the 
{\em global} net reaction\cite{b3,b3-a},
\begin{equation}\label{net-reaction}
C_3S_{(s)} + 4H_2O_{(\ell)}
\stackrel{k}{\rightharpoonup} C_{1.5}SH_{2.5(s)} + 1.5  CH_{(s)},
\end{equation}
where $k$ denotes an effective rate constant characterizing the 
solid to solid conversion rate.
Because Eq.~(\ref{net-reaction}) describes a net reaction   
originating from various subprocesses the 
stoichiometric numbers in Eq.~(\ref{net-reaction})
are {\em not} related to the kinetic exponents of 
the overall reaction rate law. 
The kinetics of cement hydration has been experimentally studied to
some extent by calorimetric and conductimetric measurements\cite{b3,b1},
by X-ray diffraction\cite{b31,b32}, by 
Raman spectroscopy\cite{b33}, 
as well as by NMR spectroscopy\cite{b19,b20,b21,b22}.

\subsection{NMR Measurements}\label{NMR_Measurements}
$^{29}Si$ NMR spectroscopy has been used to experimentally determine 
the $CSH$ growth rate\cite{b19,b20,b21,b22}. 
The employed method (Single Pulse Excitation
with Magic Angle Spinning (MAS/SPE)) is based on the fact that in the NMR 
spectrum the signal of the $C_3S$ silicon nuclei (monomer)
is separated from those of 
$CSH$ silicon nuclei (dimers and trimers)\cite{b18}. 
The distinction of dimer (silicon chain ends) 
and trimer (inside silicon chains) signals
has been used to investigate the polymerization process within 
the cement hydrate, see for example Ref.~\ref{b20}. 

In order to 
enhance the signal/noise ratio one generally needs to accumulate 
a great number of times the NMR signal. To achieve these 
accumulations and in order to obtain quantitative information 
about the populations of the different chemical species 
present in the specimen, one needs to strictly respect the relaxation 
of each population between subsequent accumulations. 
We have selected from the literature results obtained taking this
condition into account\cite{b19,b20,b21}. NMR signal intensities, $I_k$,
are in general proportional to the absolute number $n_k$ of excited 
silicon nuclei of species $k$ within the sample, $I_k = \Gamma n_k$, with 
$\Gamma$ being the constant of the measurement (species $k$ 
denotes a silicon nuclei belonging to a monomer, dimer, or trimer
silicon cluster). 
The proportionality constant is usually eliminated by considering 
relative signal intensities $I_k/\sum_j I_j$ which correspond to 
mole fractions $x_k=n_k/\sum_j n_j$. The constant $\sum_j n_j$ 
defines the absolute silicon mass scale. 
The NMR measurements we refer to, provide plots of the time-dependent 
mole fractions for monomer, dimer and trimer silicon 
nuclei\cite{b19,b20,b21}. 
However, plots for mole fractions show,  
as will be demonstrated below, more complex algebraic behavior.

\subsection{Schematic Representation}\label{Schematic_Representation}
In order to get an idea about the global hydration kinetics of cement
we consider the chemical {\em net} reaction 
(\ref{net-reaction}) in a more schematic representation, 
\begin{equation}\label{schematic-net-reaction}
A \stackrel{k}{\rightharpoonup} B,
\end{equation}
where the species $A$ and $B$ refer to silica nuclei of $C_3S$ and 
$CSH$ respectively. 
We denote the anhydrous $C_3S$ (monomer)
mole fraction as $x_{A}(t)$ and the $CSH$ 
(dimer as well as trimer) mole 
fraction as $x_{B}(t)$, i.e., $x_{A}(t)+x_{B}(t)=1$.

The reader is reminded that Eq.~(\ref{schematic-net-reaction}) describes  
precisely a solid--state transformation as mentioned in the introduction, 
however, there is one fundamental difference in that the dimeric and trimeric 
silicon nuclei $B$ are not {\em initially solvated} within the silicon 
monomers $A$. The $B$ nuclei become rather formed within the liquid solution 
which does not appear in Eq.~(\ref{schematic-net-reaction}).
 
It is here the place to say something about the range of validity, 
the similarities, and the differences of 
Eqs.~(\ref{net-reaction}) and (\ref{schematic-net-reaction}).

Strictly speaking, the above net reactions can only hold, if the employed 
water to cement weight ratio is not too high, that is to say
the number of solvated ions should be always orders of magnitudes 
lower than the number of molecules belonging to the 
solid state, i.e., Eqs.~(\ref{net-reaction}) 
and (\ref{schematic-net-reaction}) are {\em inconsistent} in the 
limit of infinite dilution.  
A proximate 
condition for this is a comparison of the cement to water concentration 
with the solubility of Portlandite, $S_{CH}$, 
$n_{C_3S}(t_0)/V_{H_2O}(t_0)\gg S_{CH}\approx 
2\cdot 10^{-2}\,mol\,liter^{-1}$, or $w/c \ll 200$. The here considered
experimental data certainly fulfill this condition, i.e.,  
$w/c \approx 0.5$. 

Furthermore Eq.~(\ref{net-reaction}) describes a solid--liquid to solid 
transformation, however, Eq.~(\ref{schematic-net-reaction}) a 
solid to solid transformation. It is thus natural that both reactions 
describe very different things, if the initial 
water to cement ratio is so low that the water severely becomes a 
limiting reactant for the hydration process (case of very thick pastes). 
To be more specific, the hydration following Eq.~(\ref{net-reaction}) 
cannot be completed for {\em chemical} reasons, 
if $4 n_{C_3S}(t_0) > n_{H_2O}(t_0)$ or equivalently 
if $w/c < (w/c)^* = 0.31$. It is, a priori, difficult to judge whether
the employed experimental water/cement ratio $w/c=0.5$ 
is sufficiently large 
in order to exclude a systematic effect 
on hydration due to limiting water. However, as we will see further below,
considering the kinetics of the `deacceleration' period, 
water is very unlikely a chemically limiting reactant.  
It is, however, not possible to determine from the considered 
measurements any water-specific kinetic exponent for the associated 
rate law of Eq.~(\ref{net-reaction}). The determination of such 
a dependency, would require a set of experiments, conducted for 
different initial water to cement ratios.
Similarly it is not possible to determine kinetic exponents for
Portlandite ($CH$) from the considered measurements, because $n_{CSH}$
is directly proportional to $n_{CH}$ at every instant.
In order to isolate the kinetic influence of precipitated Portlandite 
on the overall hydration kinetics
one needs to break up the direct proportionality between $n_{CSH}$
and $n_{CH}$. Preferable this could be done by examination of the 
hydration rate dependence for different initial $CH$ admixtures, i.e.,
$n_{CH}(t_0)> 0$.
Bearing the foregoing caveats in mind, it should be obvious why we      
consider the schematic reaction (\ref{schematic-net-reaction}). 
We are only in position to characterize the kinetic influence 
of reactants and products on the cement hydration.  
But already this limited information will be, in our opinion,
informative.
 
\subsection{Kinetic Relationships}\label{Kinetic_Relationships}
In the following we consider 
the ratio of silica mole fractions,
\begin{equation}\label{f} 
f(t)=x_{B}(t)/x_{A}(t),
\end{equation}
varying between $0$ (no hydrates at time $t=0$) and $\infty$ 
(complete hydration at time $t=\infty$).    
Figure \ref{fig1} shows a double-logarithmic 
plot of $f(t)$ versus the hydration time. 
This figure reveals several interesting facts about the hydration 
kinetics which will be addressed in this paper. 
We note that the considered 
experimental data have been taken from three 
independent publications obeying similar 
experimental conditions  
(room temperature, normal pressure, geometry of specimens, and initial
water/cement weight ratio $w/c=0.5$), 
see \cite{b19,b20,b21} for details. 
The relatively 
narrow scatter of the data over order of magnitudes demonstrates 
consistent and reproducible measurements. 

We observe, within in the scatter, two power laws separated by 
a sharply defined characteristic time 
$t_\times=16\,hours$ with value $f_\times \equiv f(t_\times)=0.6$.
For the `early' 
hydration period ($t<t_\times$) we obtain, 
\begin{equation}\label{early-powerlaw}
f(t)=f_\times\, (\frac{t}{t_\times})^\gamma,
\end{equation}
with an exponent $\gamma =2.5$. 
Below 5 hours of hydration time there are no 
quantitative NMR data available, because of the bad signal/noise ratio
for the $CSH$.
Above $t_{\times}=16\,hours$  
the hydration dramatically slows down,
\begin{equation}\label{later-powerlaw}
f(t)=f_\times\, (\frac{t}{t_\times})^\delta, 
\end{equation}
with $\delta = 0.5$. 
This second period takes place from $16\,hours$ to 
at least $8\cdot 10^3\,hours$ of hydration time.
Equations (\ref{early-powerlaw}) and (\ref{later-powerlaw}) 
define a continuous dependency in time. However, the first  
derivative of $f(t)$ taken at $t_\times$ is not continuous, which might 
appear unphysical. The discontinuity in $df/dt$ arises, because we 
have assumed a zero crossover range in Eqs.~(\ref{early-powerlaw}) and
(\ref{later-powerlaw}) to obtain the best representation of 
experimental data. It is interesting to note, that an ansatz of the 
form 
\begin{equation}\label{ft-fit}
t/t_\times =2^{-1/m}[(f/f_\times)^{\alpha m}
+ (f/f_\times)^{\beta m}]^{1/m},
\end{equation} 
with proper chosen exponents $\alpha=0.4$
and $\beta=2.0$
gives a quite satisfactory representation of data for high integers of
$m$, see Fig.~\ref{fig1}\cite{com3}.
The best representation is found for $m\to\infty$, 
which corresponds to Eqs.~(\ref{early-powerlaw}) and 
(\ref{later-powerlaw}).

In the following we propose relations between  
the above mentioned exponents $\gamma$ and $\delta$ and the 
kinetic exponents of an overall rate equation. The observed exponents
will allow for some general statements about the hydration mechanisms.

We note, that from Eq.~(\ref{f}) the trivial relations
\begin{equation}\label{e1}
x_{A}(t)=\frac{n_{A}(t)}{n_{A}(t_0)} =\frac{1}{1+f(t)},
\end{equation}
and 
\begin{equation}\label{e2}
x_{B}(t)=\frac{n_{B}(t)}{n_{A}(t_0)} =\frac{f(t)}{1+f(t)},
\end{equation}
follow, where $n_{A}(t_0)$ characterizes the absolute silica mass
scale.  
The overall rate equation for $x_{A}$ and $x_{B}$ has 
to be first order in time,
\begin{equation}\label{e3} 
-\frac{d}{dt}x_{A}=\frac{d}{dt}x_{B}=\frac{1}{(1+f)^2}
\frac{d}{dt}f.
\end{equation}
One particular observation from Eqs.~(\ref{e1}), (\ref{e2}), 
and (\ref{e3}) is that the global hydration rate can be represented 
as a product of powers of the mole fractions
$x_{A}$ and $x_{B}$, if 
$\frac{d}{dt}f$ follows a power-law in $f$. In such case 
the reaction is of overall order two and $f(t)$ follows a 
power-law in time, except for $\frac{d}{dt}f\sim f$ which yields an 
exponential dependency in time.
We have already mentioned, considering the experimental results in
Fig.~\ref{fig1}, that $f(t)$ follows indeed two power-laws, 
Eqs.~(\ref{early-powerlaw}) and (\ref{later-powerlaw}),
seperated by a characteristic time $t_\times$. 
Thus it is possible to extract two rate laws from the measurements; one
for `early' times $t<t_\times$ and another one  for `late' times 
$t>t_\times$. 
Consider $f(t)=f_\times\cdot(t/t_\times)^\psi$ where for times
$t<t_\times$, $\psi=\gamma$ and for times $t>t_\times$, $\psi=\delta$, 
according to Eqs.~(\ref{early-powerlaw}) and (\ref{later-powerlaw}). 
The associated first order differential equation is 
$(d/dt)f=\psi t_\times^{-1} f_\times^{1/\psi} f^{(\psi -1)/\psi}$.
Inserting this into Eq.~(\ref{e3}) and resorting with respect to
factors $x_{B}$ and $x_{A}$ [Eqs.~(\ref{e2}) and (\ref{e1})] 
we obtain the explicit rate equation in terms of mole fractions,
\begin{equation}\label{e4}  
-\frac{d}{dt}x_{A}=\frac{d}{dt}x_{B}=
\psi t_\times^{-1} f_\times^{1/\psi} 
x_{B}^{\frac{\psi -1}{\psi}} 
x_{A}^{\frac{\psi +1}{\psi}},
\end{equation}
with $\psi=2.5$ for $t<t_\times$ and $\psi=0.5$ for $t>t_\times$.

Furthermore, by passing from mole fractions to concentrations,
it is possible 
to determine the approximate
effective rate constant(s) from the experimentally observed 
crossover point, the extracted exponents, and the initial 
water to cement weight ratio. 
It is convenient to consider first the initial concentration 
$[A]\vert_{t_0}=n_A(t_0) / V_0$ 
of cement in the overall specimen volume 
$V_0=V_{C_3S}(t_0)+ V_{H_2O}(t_0)$ \cite{com2},
\begin{equation}\label{e5}
\frac{1}{[A]\vert_{t_0}}=
(1+\frac{\rho_{C_3S}}{\rho_{H_2O}}\cdot\frac{w}{c})v_{C_3S}
\approx 0.18\,liter\,mol^{-1},
\end{equation}
with $\rho_{C_3S}/\rho_{H_2O}=3.21$ and 
$v_{C_3S}=7.2\cdot 10^{-2}\,liter\,mol^{-1}$ being the relative
density and the molecular volume of cement respectively. The considered 
water to cement weight ratio is $w/c=0.5$.

We find for the `accelerated' period 
($t<t_\times$),   
\begin{equation}\label{accelerated}
-\frac{d}{dt}[A]= \frac{d}{dt}[B]=k_{t<t_\times}\cdot
[B]^{0.6} [A]^{1.4},
\end{equation}
with $k_{t<t_\times}=2.5 [A]\vert_{t_0}^{-1} t_\times^{-1}f_\times^{0.4}
\approx 7\cdot 10^{-6}\,liter\,mol^{-1}\,s^{-1}$, and for the 
`deaccelerated' period ($t>t_\times$)
\begin{equation}\label{deaccelerated}
-\frac{d}{dt}[A]= \frac{d}{dt}[B]=k_{t>t_\times}\cdot
[B]^{-1.0} [A]^{3.0},
\end{equation}
with $k_{t>t_\times}=0.5 [A]\vert_{t_0}^{-1} t_\times^{-1}f_\times^{2.0}
\approx 6\cdot 10^{-7}\,liter\,mol^{-1}\,s^{-1}$. 

\section{Interpretation}\label{Interpretation}
Before we are going to present a possible explanation for the hydration 
kinetics as manifested in Eqs.~(\ref{accelerated}) and 
(\ref{deaccelerated}) we would like to give a brief account on 
another representation of the hydration kinetics being more 
widespread in cement literature, i.e., the degree of hydration 
$\alpha (t)$. 
The degree of hydration is usually referred as the relative amount 
(mole fraction) of hydrated cement\cite{b0}, 
$\alpha (t)\equiv x_B(t)= 1 - x_A(t)$.
Therefore Eq.~(\ref{e2}) gives the relationship between $\alpha (t)$  
and $f(t)$ as defined in Eq.~(\ref{f}).
Typical experimental curves for the degree of hydration exhibit 
sigmoidal shapes in linear representations. The observed inflection 
points, however, do in general not posses any particular  
significance for a change in chemical mechanism. 
This can be seen for example
by assuming in Eq.~(\ref{e4}) $\Psi > 1$ for {\em all times}. 
The degree of hydration will show an inflection point though 
there is only a single chemical mechanism (rate law) operative. 
Therefore it is not reliable to read off characteristic times from 
$\alpha (t)$ diagrams. While $\alpha (t)$ approximates $f(t)$ well
if $f \ll 1$ (early hydration) the discrepancy becomes very 
large for $f \gg 1$ (late hydration). 

The foregoing remark has hopefully illustrated why we have avoided 
degree of hydration diagrams in our considerations. 
As another motivation for our approach we present in 
Fig.~\ref{fig_avrami} a so-called `Avrami--Plot' for the hydrated 
silica amount, i.e., a test 
on the stretched exponential relationship $x_B = 1-\,exp(-(t/\tau)^k)$.
Such empirical relationships are frequently found for overall 
transformations\cite{Hann75}. 
It can be clearly seen from  Fig.~\ref{fig_avrami} that the hydration 
data {\em cannot} be described by such a relation.

The kinetic equations Eqs.~(\ref{accelerated}) and
(\ref{deaccelerated}) allow to make somewhat more substantial statements 
about the global cement hydration mechanism(s). 
During the acceleration period
already existing hydrates {\em catalyse} 
the precipitation 
of new hydrates (positive kinetic exponent for the products in 
Eq.~(\ref{accelerated})). 
The growth of {\em connected} hydrate structures 
appears thus to be thermodynamically more favorable than an uncorrelated  
`through solution' precipitation mechanism. 

On the other hand in the deacceleration period ($t>t_\times$) 
the `hydrate layers' surrounding the cement grains
increasingly separate reacting anhydrous cement 
and water and 
thus hinder/block further $C_3S$ dissolution.
The hydrates are acting in the deaccelerated period
as {\em inhibitors} (negative exponent for 
the products in Eq.~(\ref{deaccelerated}). 
The inversed role of 
hydration products during the accelerated and deaccelerated periods
appears to be experimentally evident from the foregoing
considerations. 

\begin{figure}[htb]
\centerline{\psfig{file=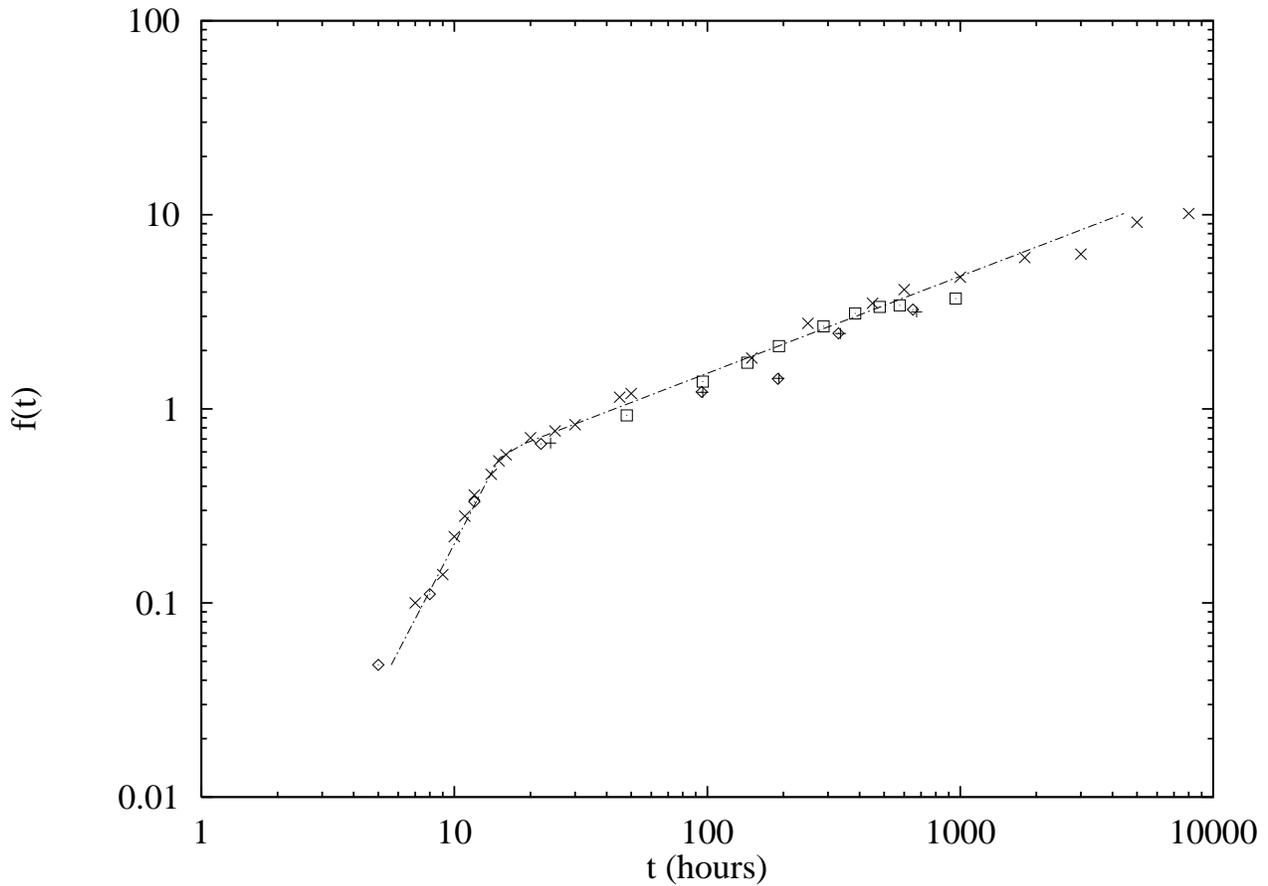,width=\textwidth}}
\vskip 5mm
\caption{Double logarithmic plot of the measured mole ratio  
between hydrated and unhydrated silica $f(t)=x_B/(1-x_B)$
  as a function of hydration time $t$ employing NMR and 
  TGA (Thermogravimetry). All measurements correspond to a water to 
  cement weight ratio $w/c=0.5$ conducted under room temperature and
  normal pressure.  
  The data are well represented by two power-laws:
  for $t<t_\times$ by Eq.~(\ref{early-powerlaw}) ($\gamma =2.5$) 
  and for $t>t_\times$ by Eq.~(\ref{later-powerlaw}) ($\delta =0.5$).
  The dashed line shows the corresponding numerical fit according to
  Eq.~(\ref{ft-fit}) employing  
  $t_\times =16\,hours$, $f(t_\times) =0.6$, and $m=20$. 
  Data from ($\diamond$) Ref.~\ref{b19} (NMR), 
  ($\times$) Ref.~\ref{b20} (NMR), 
  ($\Box$) Ref.~\ref{b21} (NMR) and ($+$) Ref.~\ref{b19} (TGA).
  }
\label{fig1}
\end{figure}

\begin{figure}[htb]
\centerline{\psfig{file=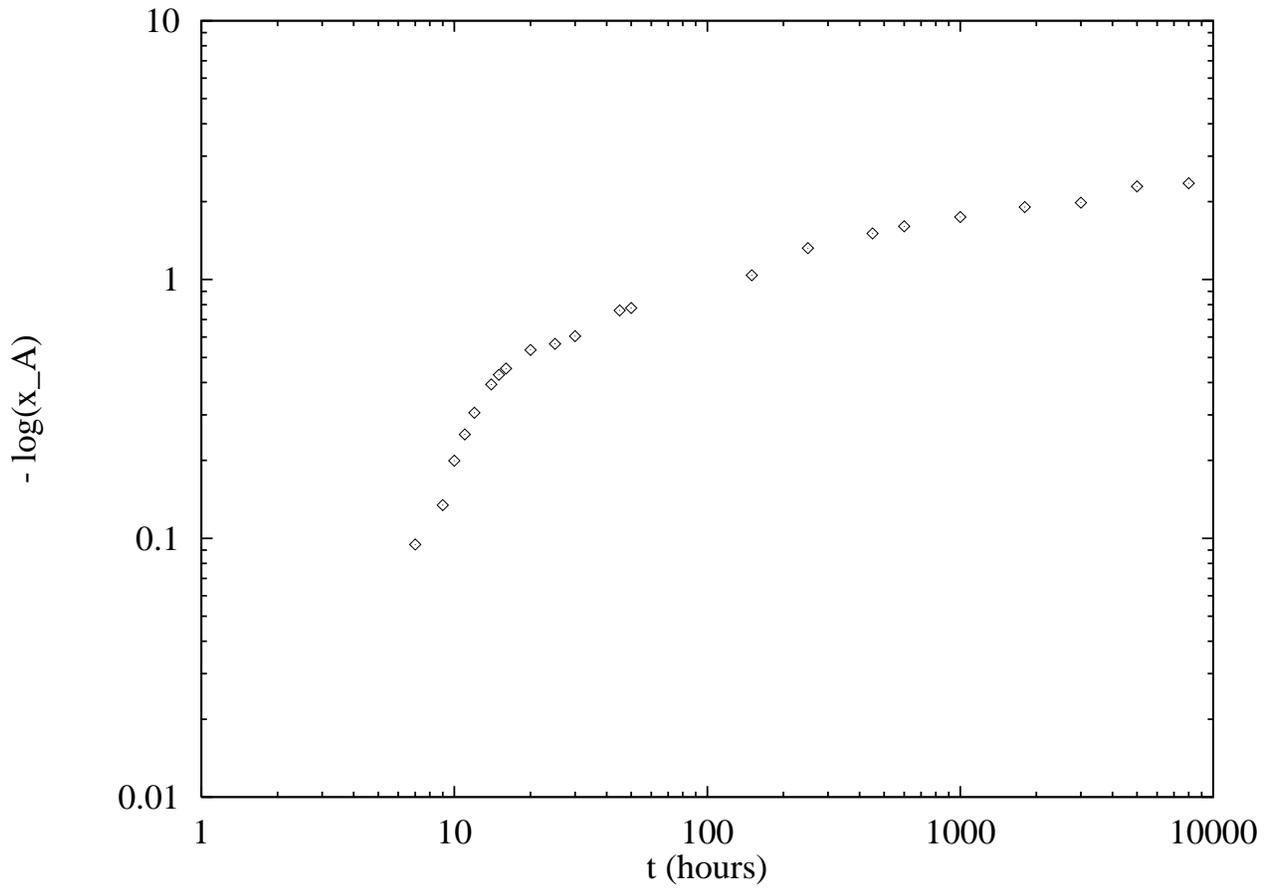,width=\textwidth}}
\vskip 5mm
\caption{`Avrami'--Plot for the amount of hydrated silica 
(double logarithmic plot of the negative logarithm of unhydrated
silica amount $-\log (x_A)$ versus time $t$). If the hydration kinetics 
would follow a generalized Avarami--Johnson--Mehl law, i.e., 
$x_A = 1-x_B = exp(-(t/\tau)^k)$, the data should lay on a {\em single}
straight line of slope $k$. It can been seen that this is {\em not} 
the case. 
The data were taken from Ref.~\ref{b20} (NMR).
}
\label{fig_avrami}
\end{figure}

A widespread assertion in cement literature is that `early' hydration is 
controlled by chemical kinetics whilst `late' stage kinetics is 
being diffusion controlled. We agree with the later assumption that 
ion diffusion most probably represents the rate controlling step
within the deacceleration period. This is in fact strongly indicated 
by the very low hydration rate at large times, see Fig.~\ref{fig1}. 

However, there is no direct indication that the 
acceleration kinetics is chemically limited.
Obviously Eq.~(\ref{accelerated}) {\em cannot} describe initial nucleation 
(approximately within the first $15\;min$) because 
there exist no products at this times at all ($[B]\vert_{t_0}=0$). 
On the other 
hand the early nucleation period is not accessible employing the 
here considered experimental techniques, so there is no conflict in 
interpretation. One just has to keep in mind that all experimental 
data points as well as the thereof extracted power--laws are  
beyond the nucleation period. 

After the 
first few minutes of bringing cement and water into contact, the ions
in solution rise their concentrations far beyond the equilibrium 
solubilities, without precipitating at all. This can be understood by 
viewing the process of heterogeneous nucleation as overcoming 
a thermodynamical
barrier (supersolubility \cite{b1}). Being supersaturated the hydrate
nucleation happens on the cement grain surfaces \cite{Jenn81,b0}. 
At this instant 
a more or less significant part of the solution is strongly
oversaturated with respect to the equilibrium solubility. The further
precipitation (growth) of the hydrates can thus be regarded to happen 
approximately within a spatial uniform oversaturated solution
(oversaturated interface layer).   
This is in so far of concern as the uniformity of `initial conditions'
is crucial to predict kinetic exponents for diffusion controlled 
reactions. It also supposes that the nucleation rate tends 
rapidly to zero as the hydrate microstructure further develops.

The hydrate microstructure has been experimentally classified for 
a water to cement ratio $w/c=0.47$.
Within the first $4\; hours$ foil--like microstructure precipitation 
is reported to happen radially away from the cement grains 
(typical dimensions $<0.5\;\mu m$ and $x_B < 10^{-2}$).
Thereafter ($<24\;hours$) the formation of a gelatinous layer 
sourrouding the cement grains have been observed (thickness
$\approx 0.5\;\mu m$ and $x_B(24\;h) \approx 0.3$). Also
needle-like precipitates have been found.
Finally after several days crumpled interlocking
foils are observed (comp.~Fig.~\ref{fig0}). 
It has been also reported that the precipitated 
microstructure morphologies are strongly influenced by the 
available interparticle spacings\cite{Jenn81}.
There is presently no straightforward way
to predict/calculate microstructural 
morphologies for such complex systems, however, recently developed 
heterogeneous reaction-diffusion models have received considerable 
interest in this context\cite{Tzsch96}.

The growth of a precipitate is kinematically very complex.  
The precipitated geometry (microstructure) cannot 
be predescribed in general beyond the initial conditions but
is rather a {\em result} of the associated interface dynamics.
The most simplest case is governed by the growth of spherical
precipitates of radius $R$ from an initially supersaturated solution
of concentration $\bar{c}$. Let $c=c(r)$ 
denote the particle concentration in the solvent,  
$c_{s\ell}$ the particle
concentration within the precipitate at the
interface, $c_{\ell s}(R)$ the particle 
concentration within the solvent at the
interface and $D$ the diffusion coefficient of 
particles in the solvent. One has from continuity of mass,
\begin{equation}\label{BoundaryEquation}
\frac{d\,R}{d\,t}= \frac{D}{c_{s\ell} - c_{\ell s}(R) }\cdot
\Bigl( \frac{d\,c}{d\,r} \Bigr)_{r=R}.
\end{equation}
This is the equation of motion for the interface (in spherical 
co--ordinates). It establishes the direct proportionality between 
{\em local} growth rate and particle current density at the interface.
In general $c_{\ell s}$ depends,  
as a consequence of interfacial tension,   
on the (local) interface curvature. If
the precipitate is not too small one often assumes 
the zero curvature limit, i.e., $c_{\ell s}(R)\approx 
c_{\ell s}(\infty)$.     
The most interesting point in Eq.~(\ref{BoundaryEquation}) is that 
the gradient of $c$ contains `global information' about the interface 
because $c(r)$ is the solution of the (stationary) diffusion equation,
\begin{equation}\label{LaplaceEquation}
\Delta\, c(r) = 0,
\end{equation}
obtained under boundary conditions $c(\infty)=\bar{c}$ and 
$c(R)=c_{\ell s}(\infty)$. 
The solution of 
Eqs.~(\ref{BoundaryEquation}) and (\ref{LaplaceEquation}) is the
well known parabolic growth law\cite{Hann75},
\begin{equation}\label{ParabolicGrowthLaw}
R^2 - R_0^2 = 2D \frac{\bar{c}- c_{\ell s}(\infty) }
{c_{s\ell} -c_{\ell s}(\infty) }(t-t_0).
\end{equation}
Similarly a parabolic solution is also obtained for the growth 
of a flat interface.
More generally does the local growth rate dynamically 
depend on (at least) two competitive mechanisms (a) flattening of high 
curvature regions due to interfacial tension and (b) sharpening 
of these regions due to preferential diffusive growth at these
`tips'. 
Numerical 
boundary integral methods have been developed in order 
to study the associated interface dynamics and instability,  
e.~g.~see Ref.~\cite{Kessler88}. For a pertubative treatment see for
example Ref.~\cite{Lovett78}.

Despite the complexity of involved microstructural transformations 
the natural question arises whether there exists at a given time
a {\em typical microstructure} and a {\em typical mode of growth}
within the system. 

Suppose the case of a vanishing nucleation rate during
precipitate growth. If the precipitating structure grows geometrically
in form of a plate (see the above experimental classification scheme)
then the variation in mole fraction of 
precipitated phase in time is predicted theoretically for $x_B\ll 1$ as
$x_B \sim (t/\tau)^{5/2}$\cite{Hann75} with $\tau$ being a characteristic 
timescale for the growth\cite{com4}. 
The plate's rim grows at a constant rate while its thickness grows 
parabolically in time, explaining the exponent $5/2$. 
For $x_B\ll 1$ one can replace the quantity 
$f(t)$ by $x_B(t)$ in all foregoing considerations. One is lead to the
conclusion that for early foil--like hydrate growth 
$\gamma = 5/2$ and $(\psi -1)/\psi =3/5$ 
in Eqs.~(\ref{early-powerlaw}) and (\ref{e4}) respectively.
This is in agreement with the considered measurements, see Fig.~\ref{fig1}. 
 
What causes the observed slowing down of the
hydration process?
If a single foil would precipitate in a spatially infinite  
supersaturated solution there would be no obvious reason for a slowing down
of the hydration process as there exists no characteristic length
scale.
However, the here considered case of cement paste is an assembly 
of small anhydrous cement grains immersed in water. The mean free 
distance between particles
has to be considered as a typical length scale for the 
transport and for the precipitation process ($x_B \ll 1$). 
The 
size of the growing flakes {\em cannot} exceed this 
because of spatial hindrance. Hence there must exist a typical time scale
$t_\times$ at which the flakes change their mode of growth into 
thickening only. We interpret this typical time as the crossover time 
$t_\times \approx 16\;hours$ observed in Fig.~\ref{fig1}. Growth of 
flakes in the thickening only mode is expected to happen parabolically
in time for diffusion limited precipitation reactions\cite{Hann75}. 
Therefore we predict the exponents of the deaccelerated period to be 
$\delta = 1/2$ and  $(\psi -1)/\psi =-1$ in
Eqs.~(\ref{later-powerlaw}) and (\ref{e4}) respectively, in agreement
with the experimental data Fig.~\ref{fig1}.

Apparently with the above assessments we have related the 
experimentally observed kinetic exponents of the hydration products 
to microstructural information. Certainly there is no unique mapping
between kinetics and geometry, but this constitutes a complex 
question in terms of Eqs.~(\ref{BoundaryEquation}) and 
(\ref{LaplaceEquation})
to be studied in future on its own right.

\begin{figure}[htb]
\centerline{\psfig{file=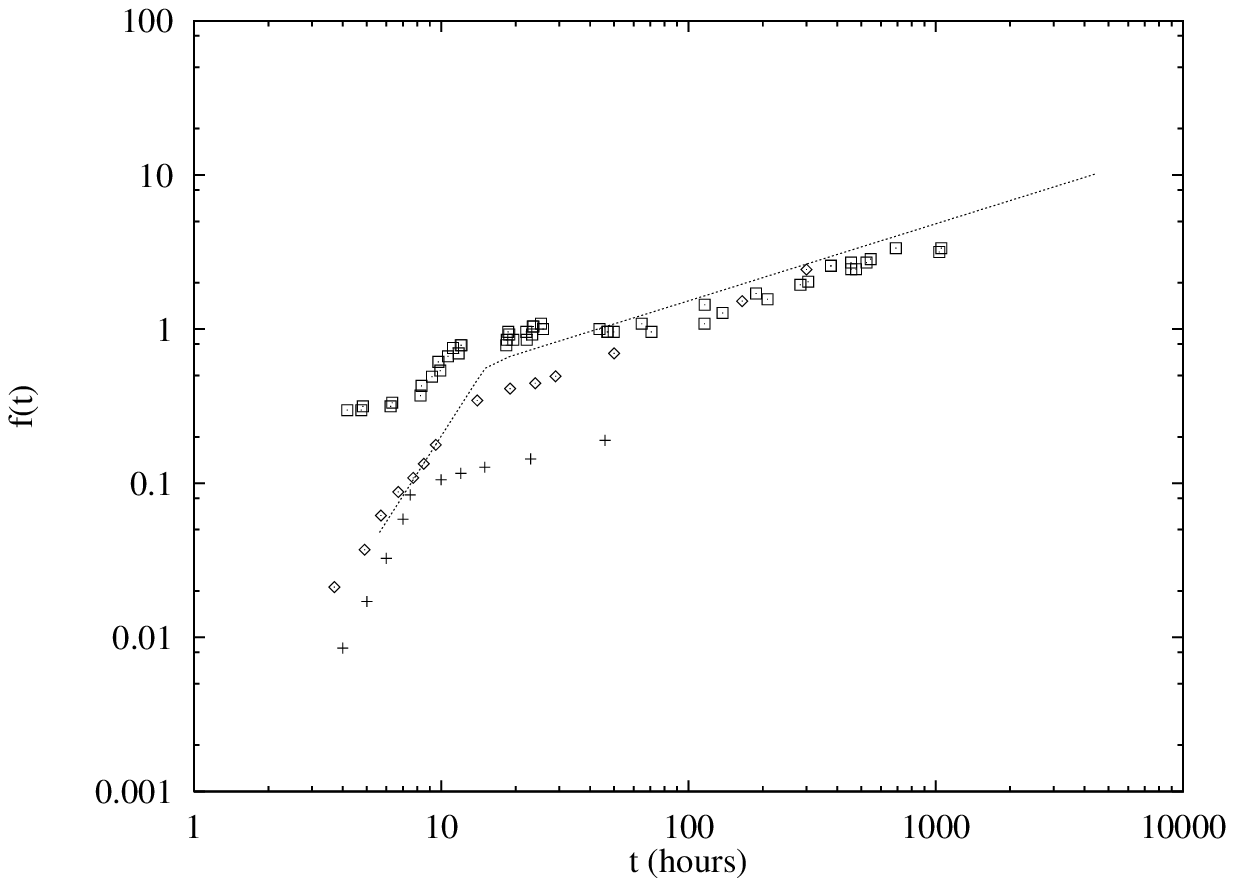,width=\textwidth}}
\vskip 5mm
\caption{Same representation as in Fig.~\ref{fig1} but for various 
water cement ratios employing XDR (x-ray diffraction) and Raman 
spectroscopy. The measurements were conducted at room temperature 
and normal pressure. The dashed line shows again the powerlaw fit
corresponding to Fig.~\ref{fig1} for comparison purposes.  
Data from 
($\Box$) Ref.~\ref{b33} (Raman, $w/c=0.4$), 
($\diamond$) Ref.~\ref{b31} (XDR, $w/c=0.45$)
and 
($+$) Ref.~\ref{b32} (XDR, $w/c=0.7$).
Note that the data do not collapse, however, for large times they 
yield similar slopes $\approx 1/2$ compared to Fig.~\ref{fig1}.
  }
\label{fig2}
\end{figure}

So far we have restricted our considerations to the case of a 
water to cement weight ratio $w/c =0.5$. 
The question arises 
whether the kinetic exponents are universal and 
how the observed typical crossover quantities $t_\times$
and $f_\times$ do depend on $w/c$. We show in Fig.~\ref{fig2} 
experimental (replotted) data for three different $w/c$ ratios. 
The data do not collapse, which is not so surprising because the 
typical interparticle spacing does depend on $w/c$. 
Qualitatively
higher $w/c$ ratios correspond to lower charateristic times $t_\times$
and lower hydrate `amounts' $f_\times$. For early times we observe 
kinetic exponents that do (apparently) depend on $w/c$. 
Interestingly are the kinetic exponents $\delta$ for the deacceleration 
period in all cases close to $1/2$.    
However, we have not studied this in further
detail because there are less experimental data available than for 
the standard case $w/c =0.5$, i.e., the hydration curves need to be 
experimentally reproducible for given $w/c$.

\section{Conclusion}\label{Conclusion}
We have reconsiderd NMR measurements on the overall
hydration kinetics of tricalcium silicate pastes ($w/c=0.5$)
In Sec.~\ref{Schematic_Representation} we briefly discussed the
conditions for $w/c$ to be fulfilled in order to allow a meaningfull 
discussion in terms of a global net reaction and its schematic 
counterpart.
In Sec.~\ref{Kinetic_Relationships} we demonstrated
that the time dependent 
ratio of hydrated and unhydrated silica mole numbers can be 
well characterized by two power--laws 
in time, $x/(1-x)\sim (t/t_\times)^\psi$. For early times 
$t < t_\times$ we found an `accelerated' hydration 
($\psi = 5/2$) and for later times $t> t_\times$ 
a `deaccelerated' behavior ($\psi = 1/2$). The crossover 
time has been estimated as $t_\times \approx 16\;hours$.
We interpreted these results in terms of a global second order rate equation
indicating that (a) hydrates do catalyse the hydration
process for $t<t_\times$, (b) they do inhibit hydration for
$t>t_\times$ and (c) the value of the associated second order rate 
constant is of magnitude 
$6\cdot 10^{-7}\;-\;7\cdot 10^{-6}\;liter\;mol^{-1}\;s^{-1}$.
We have argued in Sec.~\ref{Interpretation}, by 
considering the hydration process actually being furnished as 
a diffusion limited precipitation that the exponents 
$\psi =5/2$ and $\psi =1/2$ directly indicate a preferentially 
`leaf' like hydrate microstructure. 
This argument was supported by  
experimental observations of cellular hydrate microstructures for
this class of materials.

\section*{Acknowledgments}
F.~T.~ would like to acknowledge financial support from
CEC under grant number ERBFMBICT 950009.

\end{multicols}
\vfill\eject
%
%

\begin{thebibliography}{10}

\bibitem{Hann75},\label{Hann75}
  V. Raghavan and M. Cohen in {\em Treatise on Solid State Chemistry, 
  Vol.~5 Changes of State}, ed. by N.B. Hannay (Plenum Press, London, 
  1975) pp.~67--127.	

\bibitem{Saut96}\label{Saut96}
  G. Sauthoff, J. de Physique IV {\bf 6}, 87-97 (1996).

\bibitem{Oelk96}\label{Oelk96}
  E.H. Oelkers and P.A. Bjorkum, American Journal of Science {\bf 296}, 
  420-52 (1996).

\bibitem{c1}\label{c1}
  M.D. Cohen and R.D. Cohen, J. Mat. Sci. {\bf 23}, 3816-20, (1988).

\bibitem{b3}\label{b3}
  P. Barret and D. Bertrandie, Journal de Chimie Physique {\bf 83}, 
  11/12, 765-75, (1986).

\bibitem{b3-a}\label{b3-a}
  This is an approximation. A particularity of the cement hydration 
  problem is that the appearing cement hydrate exhibits a variable
  stoichiometry in course of its formation (`solid solution').  
  The selected stoichiometries in Eq.~(\ref{net-reaction}) 
  correspond to the `late stage' proportions in cement 
  hydration\cite{b3}.
 
\bibitem{b1}\label{b1}
  D. Damidot and A. Nonat, Advances in Cement Research {\bf 6} (21), 
  27-35, (1994).

\bibitem{b31}\label{b31}
  S. Tsumura, Zement Kalk Gips {\bf 11}, 511-8 (1966).

\bibitem{b32}\label{b32}
  I. Odler and H. Doerr, Cement and Concrete Research {\bf 9}, 
  (2), 239-48 (1979).

\bibitem{b33}\label{b33}
  M. Tarrida, M. Madon, B. Le Rolland and P. Colombet, 
  Advn. Cem. Bas. Mat. {\bf 2}, 15-20 (1995).

\bibitem{b19}\label{b19}
  C.M. Dobson, D.G.C. Goberdhan, J.D.F. Ramsay and S.A. Rodger, 
  J. Mat. Sci. {\bf 23}, 4108-14 (1988).

\bibitem{b20}\label{b20}
  A.R. Brough, C.M. Dobson, I.G. Richardson and G.W. Groves, 
  J. Mat. Sci. {\bf 29}, 3926-40 (1994).

\bibitem{b21}\label{b21}
  J. Hjorth, J. Skibsted and H.J. Jakobsen,
  Cement and Concrete Research {\bf 18}, 789-98 (1988).

\bibitem{b22}\label{b22}
  S.U. Al-Dulaijan, G. Parry-Jones, A.-H. J. Al-Tayyib and
  A. I. Al-Mana,
  J. Am. Ceram. Soc. {\bf 73}, (3), 736-39 (1990);
  H. Justnes, I. Meland, O.J. Bjoergum and J. Krane, 
  Adv. Cem. Res. 3, (11), 111-16 (1990).

\bibitem{b18}\label{b18}
  J.G. Engelhardt and D. Michel, {\em High-Resolution solid state 
  NMR of silicates and zerlites} (J. Wiley Editions 1987).

\bibitem{com3}\label{com3}
Note that $f(t)$ cannot be expressed as the sum of two powers in time 
because $\gamma >\delta$, compare Eqs.~(\ref{early-powerlaw}) and
(\ref{later-powerlaw}). However, the time can be expressed as a sum 
of two powers of $f(t)$.  

\bibitem{com2}\label{com2}
We assume here a constant and characteristic reference volume $V_0$.
This is an approximation in that the hydrating system 
Eq.~(\ref{net-reaction}) undergoes a chemical shrinkage in course 
of the hydration (nominal $\approx 10\,Vol\,\%$, however, in practice 
less).

\bibitem{b0}\label{b0}
  H.F.W. Taylor, {\em Cement Chemistry}, (Academic Press, London, 
  1990).
 
\bibitem{Jenn81}\label{Jenn81}
H.~M.~Jennings, B.~J.~Dalgleish and P.~L.~Pratt, J. Am. Ceram. Soc. 
{\bf 64}, (10), 567--72 (1981).

\bibitem{Tzsch96}\label{Tzsch96}
F.~Tzschichholz, H.~J.~Herrmann and H.~Zanni, Phys. Rev. E {\bf 53},
(3), 2629-37 (1996).

\bibitem{Kessler88}\label{Kessler88}
D.~A.~Kessler, J. Koplik and H. Levine, Adv. Phys. {\bf 37}, 
255-336 (1988).

\bibitem{Lovett78}\label{Lovett78}
R.~Lovett, P.~Ortoleva and J.~Ross, J. Chem. Phys. {\bf 69}, (3), 
947 (1978).

\bibitem{com4}\label{com4}
The timescale $\tau$ arises from such quantities as 
initial oversaturation, diffusion coefficients, and water--hydrate 
interface properties. 

\end{thebibliography}
\end{document}